# Is your Schema Good Enough to Answer my Query?

Yuanwei Zhao
College of Software
Jilin University
Changchun,Jilin,China
zhaoyuanwei0410@163.com

Lan Huang
College of Computer Science and Technology
Jilin University
Changchun,Jilin,China
huanglan@jlu.edu.cn

Bo Wang
College of Computer Science and Technology
Jilin University
Changchun,Jilin,China
826704988@qq.com

Dongxu Zhang
College of Software
Jilin University
Changchun,Jilin,China
1372443644@qq.com

Rui Zhang[1]
College of Computer Science and Technology
Jilin University
Changchun,Jilin,China
rui@jlu.edu.cn

Fausto Giunchiglia
Department of Computer Science and Information Engineering (DISI)
University of Trento
Trento,Italy
fausto.giunchiglia@unitn.it

## ABSTRACT

Ontology-based data integration has been one of the practical methodologies for heterogeneous legacy database integrated service construction. However, it is neither efficient nor economical to build the cross-domain ontology on top of the schemas of each legacy database for the specific integration application than to reuse the existed ontologies. Then the question lies in whether the existed ontology is compatible with the cross-domain queries and with all the legacy systems. It is highly needed an effective criteria to evaluate the compatibility as it limits the upbound quality of the integrated services. This paper studies the semantic similarity of schemas from the aspect of properties. It provides a set of in-depth criteria, namely coverage and flexibility to evaluate the compatibility among the queries, the schemas and the existing ontology. The weights of classes are extended to make precise compatibility computation. The use of such criteria in the practical project verifies the applicability of our method.

## 1 Introduction

Vast emergence of management information systems challenges the classical database techniques to answer the query across multiple heterogeneous legacy database systems. Mediator serves as a middleware between such database systems and the unified query services. It rewrites the queries and unites the answers according to the integrated schemas of those databases. Then the unavoidable problem arises as how to choose the candidate databases to fulfill the cross-system queries?

Suppose we are querying the profile of a student from the multiple legacy information systems in the smart university scenario: with the grades from the academic affairs office, the activities from the student union, the reading and thesis from the library, the physical records from the hospital, etc. The classical data integration solution is to map the universal query to the local queries for each data source, and reformat the answers into the form of the universal query. This solution relies heavily on the universal ontology that describes all the schemas of the data sources. A quick and dirty solution may lay on the cross-domain ontology construction through such systems, then build the mediator on top of such an ontology. But this is not always applicable nor economical for extension in the future. Yet a more practical way is to reuse the state-of-the-art general-purpose ontologies such as dbpedia, schema.org , etc. Then the question transforms to how to choose the existing ontologies?

Different approaches have been proposed with frameworks from the engineering point of view and achieved promising successes, but only a few has touched the evaluation phase of this problem. Even in the few researches, majority are qualitative discussions rather than computational criteria.

[1] Corresponding Author: Rui Zhang

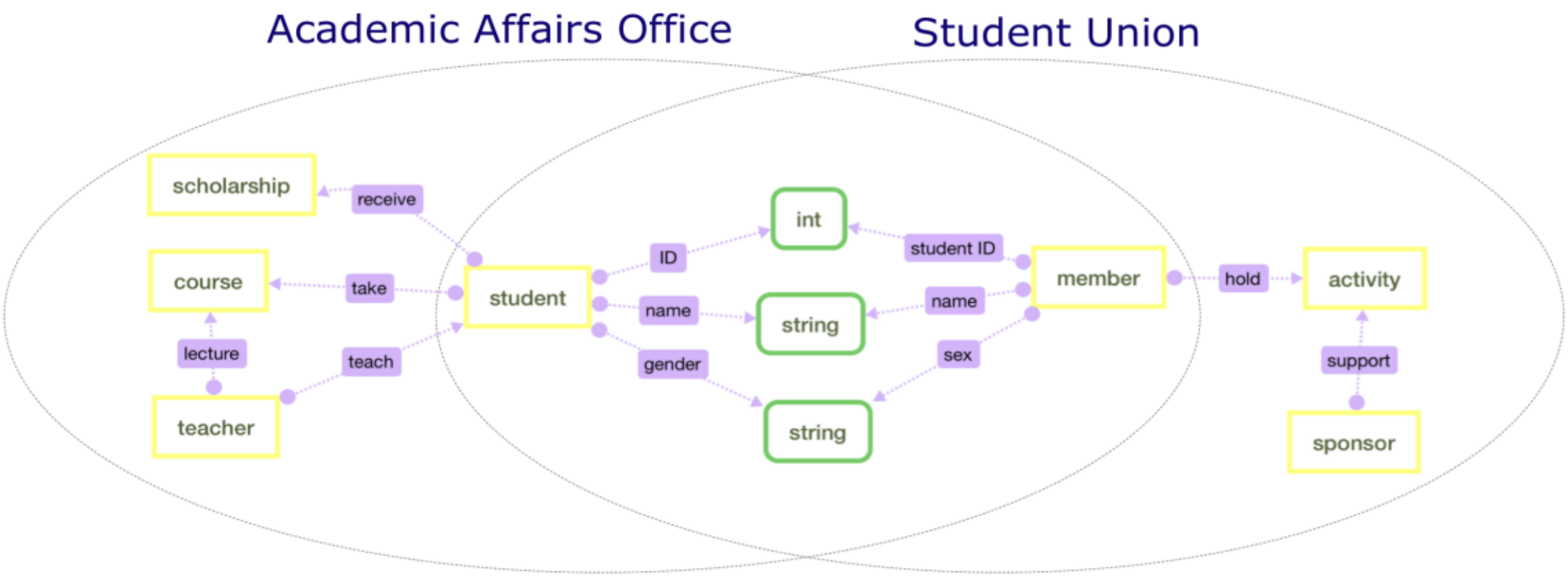

**Figure 1: Equivalent etype in different SKG**

This work roots in the European project `ISCF HDRUK DIH Sprint Exemplar: Graph-Based Data Federation for Healthcare Data Science (https://gtr.ukri.org/projects?ref=MC_PC_18029)' which aims at a solution on the medical domain heterogeneous legacy system integration between Northern Ireland and Italy. During the project, we faced the problem of semantic diversity on language level and knowledge level[1]. There are mismatches between the terms in the two natural languages, i.e. English and Italian even in the same domain on physician. Language-level diversity are solved by general-purpose dictionaries relatively straight-forward. The knowledge-level diversity is defined as the many-to-many mappings between the entity types and the properties[2].

To neglect the obscure professional terms in the medical domain, we use the student profile scenario shown in Figure 1 to motivate our criteria. There is a table *student* in the database of the academic affairs office and another table *member* in the database of the student union. The student *Mary* has records in both. Simply language-level disambiguation does not fix this kind of mismatches. A general-sense ontology will never assume this semantic equivelance either. Therefore, it is needed for an effective criteria on the knowledge level to verify this semantic similarity on the overlaps of the two schemas. This paper proposes a property based semantic similarity criteria to evaluate the overlaps of schemas. By property, we mean the attribute in the database schema to represent the entity type. For example, *birth-date* and *supervised-by*, in the table of *student* is to represent the entities such as *Mary* as a type (in a database table). Such properties form the feature vector of an entity type. The vector is first disambiguated by the language-level tools and then matched according to the semantic similarity in three-folds: label, property and individual. The thresholds for each fold are dependent on the *smart university* domain and helped to achieve the optimal precision. Then the properties are propagated through the *is-a* hierarchy of the entity types to accumulate the weights of those types. Such weights are used to calculate the coverage and flexibility of the schema overlaps. It is rational to reuse the entity types in the corresponding schemas (of the data sources or the general-sense ontology), when the overall semantic similarty exceeds the predefined domain-specific threshold.

With the criteria, the overlaps of schemas are quantified. One can choose an exsting ontology to reuse for certain competency queries accordingly rather than to imagine qualitatively. This criteria will enhance the methodology for knowledge and data integration and help to reuse ontologies for various domains.

The following of the paper is organized as: In the second section, we will introduce the related concepts. In section 3, we will give the measurement method of semantic equivalence and the formula of quantitative calculation. In the fourth section, we will give the definition of weight and the calculation method of two evaluation indexes based on weight and we will carry out experiments in section 5 to verify the correctness and effectiveness of the proposed method. The related work will be introduced in Section 6. The last section is a summary.

## 2 Problem Definition

In this part, we mainly introduce the concepts of SKG and two evaluation criteria, including coverage, flexibility and their definition.According to the abstract level of knowledge, knowledge graph can be subdivided into schema knowledge graph (SKG) and data knowledge graph (DKG). The former is the knowledge structure at the level of concepts and attributes, while the latter is the data structure at the level of individuals. Figure 1 shows two different SKGs in the same scenario. DKG is difficult to reuse because of the structural differences, but SKG can be reused in many general fields. Schema.org, DBpedia [3]contains a large number of high-quality SKGs with uniform specifications. When constructing a new knowledge graph, effectively reusing these existing high-quality

SKGs can save energy and avoid conflicts between new SKG and existing SKG.

Coverage and flexibility are two evaluation indicators used to measure the overlaps of SKGs. Coverage focuses on measuring the degree of overlap of one SKG and another, while flexibility focuses on measuring the degree of redundancy after one SKG covers another. From the semantic definition point of view, these two standards have directions.. In the experiment of section 5, we will use these two criteria to measure the matching effect between competency query (CQ)[4], schema knowledge graph(SKG) and datasets.

At present, the definition methods of coverage and flexibility mainly use the idea of set theory. As shown in Figure 2, when we need to measure the coverage rate from SKG A to SKG B, that is, $coverage(A,B)$, which will be abbreviated as $Cov(A,B)$. Then it is necessary to calculate the ratio of the uncovered part of SKG B to the whole part of SKG B. If we need to measure the redundancy of SKG A after covering SKG B, that is, $flexibility(A,B)$ ,which will be abbreviated as $Flx(A,B)$. Then it is necessary to calculate the ratio of the part of SKG A that does not cover SKG B to SKG A. It can be seen that there is no symmetry between the two independent variables in this relation,$Cov(A,B) \neq Cov(B,A), Flx(A,B) \neq Flx(B,A)$.

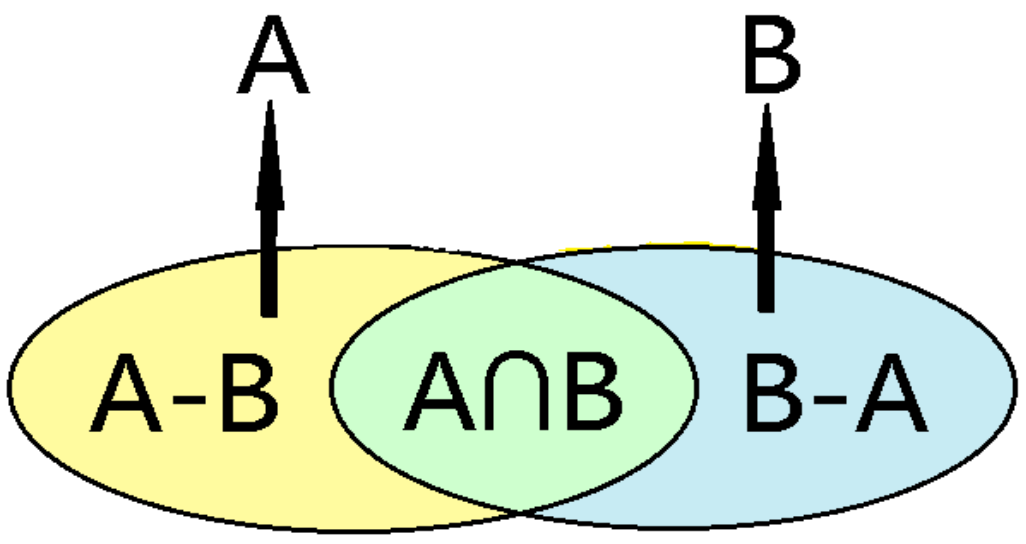

**Figure 2: Coverage and Flexibility defined in the form of set**

$$Cov(A,B) = \frac{A \cap B}{B} \tag{1}$$

$$Flx(A,B) = \frac{A - B}{A} \tag{2}$$

In the formula(1), "$A \cap B$" represents the same part of A and B and "$A - B$" in formula(2) represents the part of A that is different from B, that is, the part of A not covered by B.

## 3 Judgment of Semantic Equivalence

Taking the calculation of coverage as an example, we can define $Cov(A,B) = A \cap B/B$ with the idea of set theory. However, when we need to quantify $Cov(A,B)$, it is obviously powerless to define it with the idea of set. In this case, we need to consider how to reasonably transform the set definition into a formula which can be calculated quantitatively. In the strict set theory, it refers to exactly the same parts of A and B. However, if we regard SKG A and SKG B as a whole respectively, it is obviously impossible for them to be identical.In this case, we can consider the etype level in SKG. We can think about what are the same etypes in SKG A and SKG B, which involves how to determine the equivalence of the two etypes.

The equivalence of etypes is related to three aspects. The first is the similarity at label level, such as polysemy, can be calculated by similarity at language level. If the similarity between two etypes at label level reaches a prescribed threshold, they can be directly considered as equivalent. Secondly, there are similarities in property level, such as etype *student* and etype *member* in Figure 1. Although these two etypes are in different databases, they have very similar properties, which shows that this two etypes are similar in property level. Similarly, we should judge the similarity of different properties at the label level, such as property *gender* and *sex*, *ID* and *student ID* in Figure 1 actually represent the same meaning. If the similarity between two etypes at the property level reaches a prescribed threshold, it can be judged that the two etypes are equivalent. Finally, the similarity at the individual level from the bottom up. Suppose *Amy* is the individual of etype *student* in the database of Academic Affairs Office and the individual of etype *student* in the database of the Student Union. When we inquire, we will find that two different etypes actually correspond to the same individual. It shows that when fusing two SKGs, the etype *student* and the etype *member* belonging to different SKGs should be fused. The following is a formal definition for judging etype equivalence:

**Definition 3.1 (Semantic Equivalence):**

Given two etypes the semantic equivalence lies in 3 folds: label, property and individual.

The semantic similarity $Sim_s$ of two etypes U, V can be calculated with the formula 3.

$$Sim_s(U,V) = \alpha Sim_L(U,V) + \beta Sim_P(U,V) + \gamma Sim_I(U,V) \tag{3}$$

where

$$Sim_L(U,V) = |\overline{U},\overline{V}|$$

That $\overline{P}$ is the vector formed by the natural language label of $P$ and $|\cdot|$ returns the semantic distance of the two vectors;

$$Sim_P(U,V) = \sum_{i,j} Sim_L(u_i,v_j)$$

that $u_i, v_j$ are property of U, V respectively;

$$Sim_I(U,V) = \sum_{i,j}^{m,n} e(u_i,v_j)/Min(m,n)$$

that $u_i, v_j$ are among the m, n instances of U, V respectively and

$$e(u,v) = \begin{cases} 1, \text{if u refers to the same individual as v,} \\ 0, \text{otherwise.} \end{cases}$$

Moreover, to compute formula 3, the three factors take effects in the following conditions.

$$(\alpha,\beta,\gamma) = \begin{cases} (1,0,0), if\ Sim_L(U,V) > T_L \\ (1,1,0), if\ Sim_L(U,V) \le T_L\ and\ Sim_P > T_P \\ (1,1,1), otherwise. \end{cases}$$

Here, $T_L, T_P$ stand for the thresholds for the similarity on the label level and property level respectively.

Furthermore, an overall threshold $T_S (T_S < T_L\ and\ T_S < T_P)$ is defined with the empirical value of the expert to approximate the semantic equivalence. E.g.,if two etypes Student in the database of the education affair office and Member in the database of the student union, that $Sim_S(Student, Member) > T_S$, the two etypes are regarded semantic equivalent.

**Theorem 3.1:** The method proposed in Definition 3.1 can measure the semantic equivalence of two etypes.

Proof: First, if two etypes $U, V$ reach a given threshold at the label level,there is

$$Sim_L(U,V) > T_L$$

then in Formula 3, $(\alpha,\beta,\gamma) = (1,0,0)$,there is

$$Sim_S(U,V) = Sim_L(U,V) > T_L$$

According to the limitation of $T_S$ ,$T_S < T_L$ ,we can deduce

$$Sim_S(U,V) > T_S$$

So we can conclude that the two etypes $U, V$ are regarded semantic equivalent.

On the other hand, when $Sim_L(U,V) < T_L$ but $Sim_P > T_P$,then $(\alpha,\beta,\gamma) = (1,1,0)$ ,there is

$$Sim_S(U,V) = Sim_L(U,V) + Sim_P(U,V)$$

$$Sim_S(U,V) > Sim_L(U,V) + T_P > T_P$$

and $T_S$ meet $T_S < T_P$, so

$$Sim_S(U,V) > T_S$$

which indicates that the two etypes $U, V$ are regarded semantic equivalent.

Finally, in the third case when $Sim_L(U,V) \le T_L\ and$ $Sim_P \le T_P$, there is $Sim_S(U,V) = Sim_L(U,V) + Sim_P(U,V) + Sim_I(U,V)$. It is obvious consistent.

In conclusion, the method proposed in Definition 3.1 can judge the semantic equivalence of two etypes. When the calculated results of two etypes at the level of label and property reach the given threshold, the two etypes can be regarded as equivalent.□

# 4 Calculation of Coverage and Flexibility

Section 3 shows how to judge the equivalence of etypes in SKG A and SKG B is given, so the semantics of set definition $A \cap B$ are several etypes in SKG A and SKG B which can be judged to be equivalent. In this case, the precondition for quantitative calculation of formulas (1) and (2) are met, and further consideration is given to how to define the weights of different etypes according to the mutual influence between etypes in a fixed SKG. Finally, based on the weight, the quantitative calculation methods of coverage and flexibility is given.

## 4.1 Weight

Before introducing the definition of weight, let's take the SKG extracted in Figure 1 (as shown in Figure 3) as an example (hereinafter referred to as SKG A).

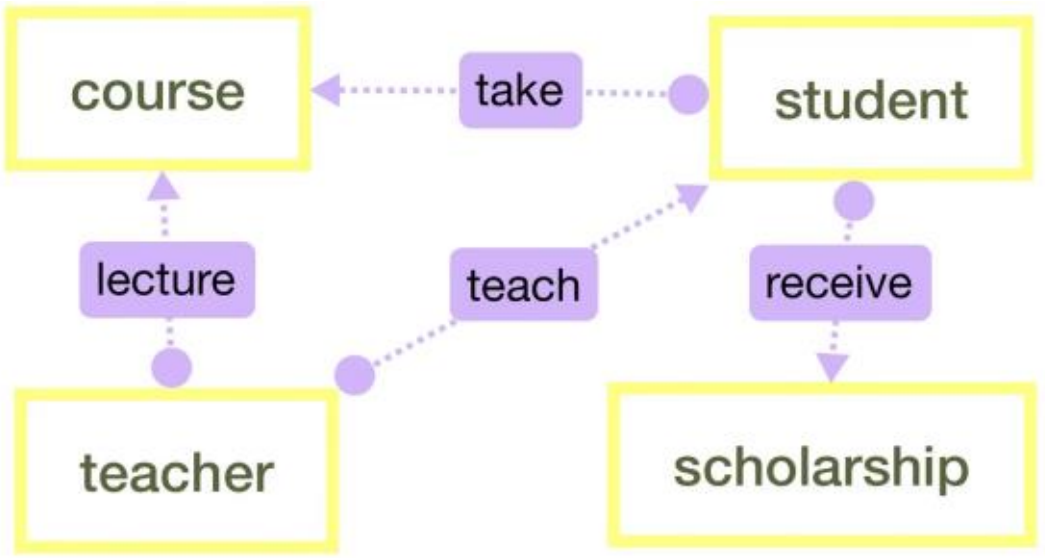

**Figure 3: SKG Extracted from Figure 1**

In this SKG A, there are four etypes: {student, teacher, course, scholarship}, and there four object properties {receive, take, teach, lecture} between these four etypes. It can be seen that among these four entities, student is associated with the other three entities, while scholarship is associated with student only. In this case, if it is necessary to evaluate the coverage effect of another SKG B on SKG A, suppose that SKG B contains etype {teacher, course, scholarship}, but etype student is missing, that is, etype student in SKG A is not covered, and the consequences brought by this lack are relatively serious. The absence of etype student will involve the lack of the connection between etype student and other etypes, which will easily lead to the incomplete and inaccurate phenomenon when two SKGs cover each other. Therefore $Cov(B, A)$ indicating the coverage degree of SKG B to SKG A and the calculation result of $Cov(B, A)$ should be smaller in this case.

On the other hand, if SKG B can't cover the scholarship only, but covers the other three etypes, it is still incomplete, but it will not affect the relationship between the other etypes. That is to say, the negative impact of integrity is relatively small, so the calculation result of $Cov(B, A)$ should be greater than the previous situation.

By the same token, when we need to calculate the redundancy degree after SKG A covers another SKG C, if there is no etype student in SKG C, that is to say, although etype student is very important in SKG A, it does not actually appear in SKG C, which means that etype Student in SKG A does not play a role.That is, SKG A has serious redundancy. $Flx(A,C)$ indicates the redundancy degree after SKG A covers SKG C,so the calculation result of $Flx(A,C)$ should be greater. On the contrary, if etype scholarship in SKG A does not participate in the coverage of SKG C, the redundancy degree will be relatively low and $Flx(A,C)$ should be smaller than the previous situation.

According to the example above, the intuition is that different etypes should be given different weights according to their importance. The importance here is understood as the degree of dependence between them and other etypes or properties in a fixed SKG, which can be quantified as the number of object property.

**Define 4.1 (Weight of Etype):**

In a fixed SKG, the weight(E) of etype E is

$$weight(E) = \frac{|L_E|}{2 * |L|} \quad (4)$$

In which $|L|$ is the number of object property p between all etype in SKG.$L_E$ is defined as follows:

$$L_E = \{p | E \in domain(p) \ or \ E \in range(p)$$

$L_E$ is the number of all object property p that satisfy the condition.

In the denominator of weight definition, $|L|$ needs to be multiplied by 2 because each object property p is actually calculated twice ( $E \in domain(p) \ or \ E \in range(p)$).This process ensures the normalization of weights, that is, the weight of each etype is less than 1 and the weight sum of all etypes is 1, $\sum_{i=1}^{n} |L_{E_i}| = 2 * |L|$ (n is the total number of etypes in SKG).

## 4.2 Handling of is-a Relationship

When calculating coverage and flexibility, we need to calculate the number of object properties. We think that different object properties have the same influence on the importance of etype, but before calculation, we need to deal with some edges specially, such as the relationship between is-a and restriction, because its semantics contains the complex relationship between superclass and subclass.

Is-a relationship can be divided into the relationship between etypes and object properties. Is-a relationship between etypes implies that subclasses enjoy the object property of their superclass. In calculation, we will transfer the object property that has influence on the superclass to subclasse. At the same time, if all subclasses of the same superclass are associated with another etype A, we think that the superclass will also be associated with etype A due to the closed world hypothesis. Based on the above analysis of is-a semantics between etype, we preprocess before weight calculation. Because is-a relationship corresponds to the semantics of superclass to subclass and subclass to superclass, we should preprocess these two directions respectively:

*4.2.1 From Superclass to Subclass*

In SKG, if the superclass has any object property $p_1$, then the subclass should have the same object property $p_1$, which does not exist explicitly in KG, so it should be considered when calculating the weight of etype. The specific method is to assign all the object properties of the superclass to each subclass etype of the superclass, which is reflected in the fact that the weight molecule of each subclass etype increase by one for every object property queried in the calculation of the superclass etype, that is, $|L| + 1$ in formula (4).

*4.2.2 From Subclass to Superclass*

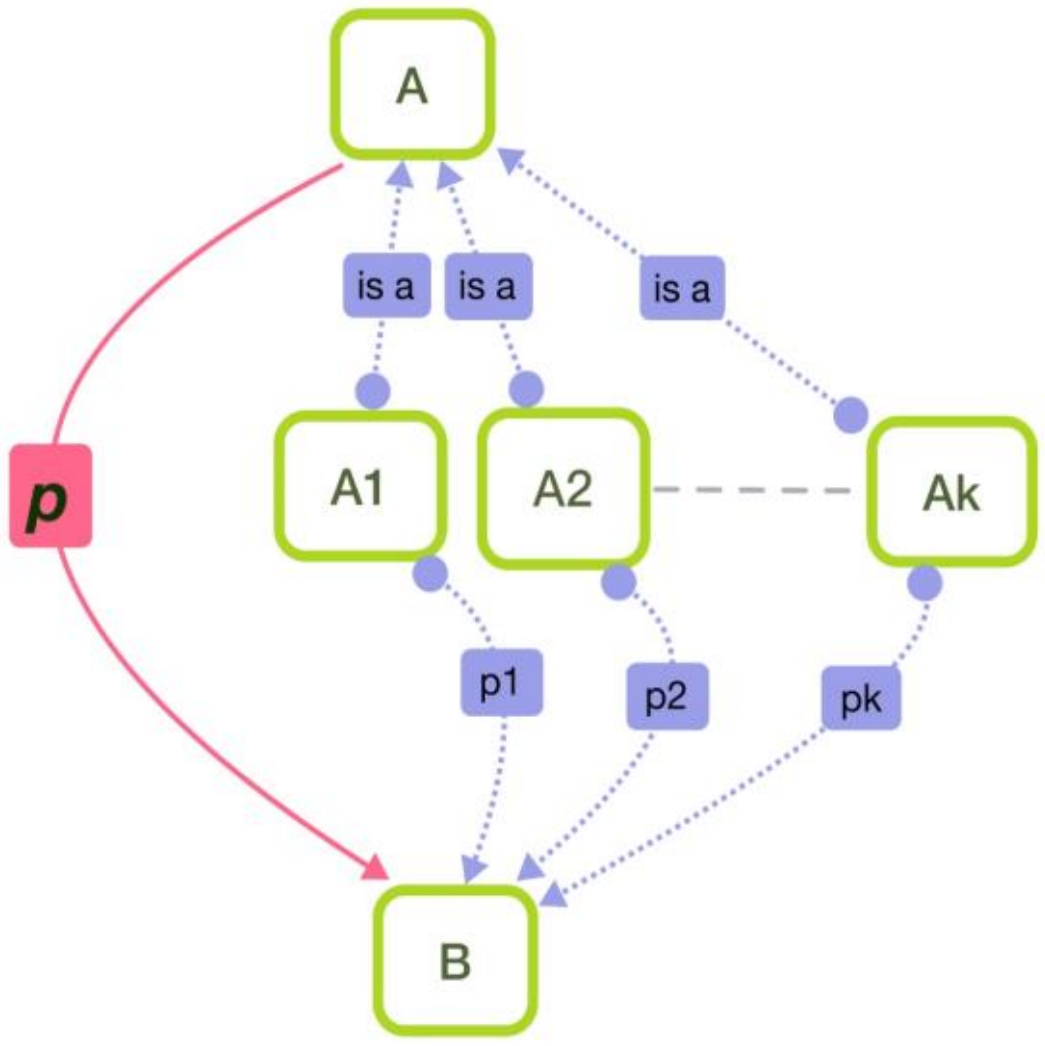

**Figure 4: Convert subclass property to superclass**

As shown in the Figure 4, it is assumed that etype $\{A_1, A_2 ... A_k\}$ are all subclasses of etype A in SKG. If all subclasses $\{A_1, A_2 ... A_k\}$ have object properties $p_i(i = 1, ..., k)$ pointing to etype B, then creates an object property $p$ pointing from etype A to etype B. Here, $p$ is only used to calculate weights, so the specific meaning of $p$ is not considered, but in theory, the common parent property of the original k edges can be extracted. In addition, we assume that these k object properties are different, otherwise they should be refined into object properties of superclasses etype A to etype B before building SKG.

On the premise of the closed-world hypothesis, that is, when we evaluate it, SKG is fixed. Therefore. On the semantic level, this operation can be explained, because if every subclass of etype A has a relationship with etype B, it means that etype A and etype B also have a certain relationship. When calculating the weight of etype A, the influence of etype B should be considered. In the specific calculation, if each subclass of etype A has an object property pointing to etype B, then the weight molecule of etype A will increase by one, that is, $|L_A| + 1$ in formula (4).

In the process of migrating object property from superclass to subclass, there are two special cases that need to be dealt with as follows.

As shown in the Figure 5 and Figure 6, suppose that in SKG, etype $A_1$ is a subclass of etype A, there is object property $p_1$ between etype A and etype B, and etype $B_1$ (only in Figure 6) is a subclass of etype B.

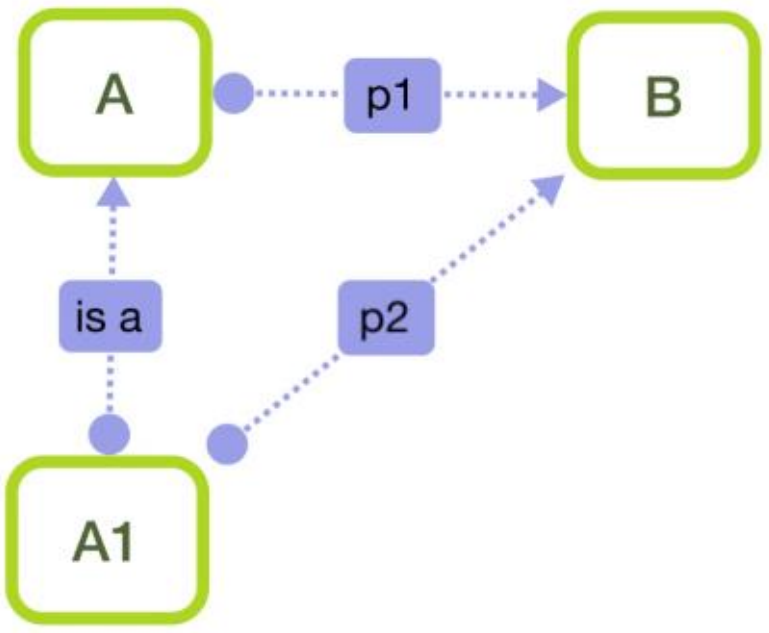

**Figure 5: Special case 1 of is-a relation processing**

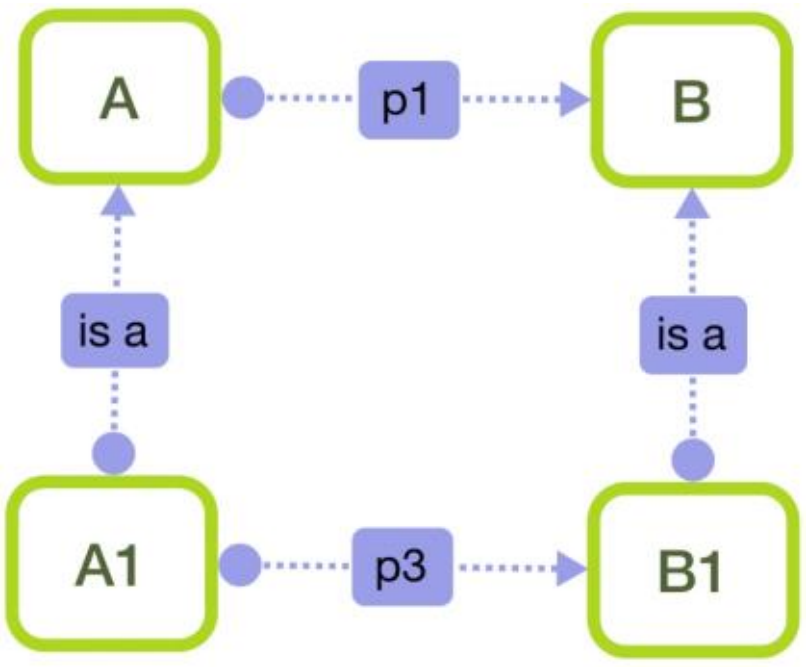

**Figure 6: Special case 2 of is-a relation processing**

1. As is shown in Figure 5, if there is an object property $p_2$ between $A_1$ and B and $p_2$ is a subclass of $p_1$, there is no need to create a new object property $p_1$ from $A_1$ to B. In other cases, create a new object property $p_1$ from $A_1$ to B.
2. As is shown in Figure 6, if there is object property $p_3$ between $A_1$ and $B_1$, and $p_3$ is a subclass of $p_3$, there is no need to create a new object property $p_1$ from $A_1$ to $B_1$, but create a new object property $p_3$ from $A_1$ to $B_1$in other cases.

After preprocessing, the is-a relationship in the original SKG is removed, and then the weight are calculated according to 4.1 .

*4.2.3 Handling of Restriction Relationship*

The owl file [5] will be one of the input files in the experiment in section 5. Protégé-OWL API[6] makes a clear distinction between named class and anonymous class. Named classes are used to create instances, and anonymous classes are used to explain in detail the logical characteristics of named classes. We can classify all individuals with the same attributes into one anonymous class, which is called Restriction[7].All types of restrictions actually describing an anonymous class that contains some individuals.When we set restrictions on a named class, we are actually describing the anonymous superclass of that class.Owl [8] defines anonymous classes through owl:Restriction, which is defined by attribute constraints applied to anonymous classes, and describes the concept of constructors in logic correspondingly through existence constraints, full name constraints and cardinality constraints of relationships.

Give a simple example: teacher, an etype, has a limit of "teaching people",so theoretically speaking, "teaching people" is a superclass of teacher. The next question is how to measure the influence of such an anonymous superclass on the weight of its subclasses. Our approach method does not consider the weight of anonymous superclass, because its significance is only to restrict subclasses. On the other hand, the object properties between anonymous superclass and other etype are still inherited by subclasses according to the method in Section 4.2.1 .

## 4.3 Calculation of Coverage and Flexibility

After defining the weights, we can define the calculation formulas of coverage and flexibility:

**Definition 4.2 (Coverage and Flexibility):**

Given a SKG X,

$$X = \{A_1: k_1, \dots, A_n: k_n, B_1: k_{n+1}, \dots, B_m: k_{n+m}\}$$

$A_i: k_i$ means that for etype $A_i$ in SKG X, $k_i = weight(A_i)$ , and a SKG Y,

$$Y = \{A_1: j_1, \dots, A_n: j_n, C_1: j_{n+1}, \dots, C_l: j_{n+l}\}$$

The coverage and flexibility of SKG X to SKG Y is given by:

$$Cov(X, Y) = \sum_{i=1}^{n} j_i \quad (5)$$

$$Flx(X, Y) = \sum_{i=n+1}^{n+l} k_i \quad (6)$$

Here are some explanations of the formula:

1. The semantic equivalent part etype between SKG X and SKG Y is $A_1 - A_n$.
2. The positions of the two independent variables in (5) and (6) cannot be exchanged, that is, the independent variables do not have symmetry $Cov(X, Y) \neq Cov(Y, X). Flx(X, Y) \neq Flx(Y, X)$ 。

**Theorem 4.1:** Formulas (5) and (6) conform to the semantics set definition (1) and (2) that is, the calculation method proposed in definition 4.1 is correct.

Proof: $Cov(X, Y)$ is defined as

$$Cov(X,Y) = \frac{X \cap Y}{Y}$$

Combined with definition 4.1, for each etype $E_i$ in SKG Y,there is

$$weight(E_i) = \frac{|L_{E_i}|}{2 * |L_Y|}$$

Where $|L_{E_i}|$ is the number of all object properties with etype $E_i$ as the starting point or end point, and $|L_Y|$ is the number of object properties in all SKG Y, defined by definition 4.2,there is

$$j_i = weight(E_i) = \frac{|L_{E_i}|}{2 * |L_Y|}$$

thus

$$Cov(X,Y) = \sum_{i=1}^{n} j_i = \sum_{i=1}^{n} \frac{|L_{E_i}|}{2 * |L_Y|} = \frac{\sum_{i=1}^{n} |L_{E_i}|}{2 * |L_Y|}$$

$X \cap Y$ is the same part of two SKGs, namely etype $A_1 - A_n$, and the number of object properties of this part is $\sum_{i=1}^{n} |L_{E_i}|$. On the other hand, the total number of object properties of SKG Y is $|L_Y|$, as shown in Figure 3, each object property links two etypes, and both etypes count this object property when calculating the weight. Therefore, each object property is actually counted twice, so $2 * |L_Y|$ represents the times of all object properties participating in statistics in SKG Y.

To sum up $\frac{\sum_{i=1}^{n} |L_{e_i}|}{2 * |L_Y|}$ represent the meaning of $\frac{X \cap Y}{Y}$, the calculation method of definition 4.2 is correct. The correctness proof of $Flx(X,Y)$ can be obtained in the same way.□

## 5. Experiment

### 5.1 Data and Methodology

The experiment is on a desktop pc with processor Intel i7-6700hq, 8GB DDR4 1333 Ram, and SSD hard disk 900GB. To make the experiment neutral, we used business domain data, including 3 types of schemas describing the: competency query (CQ), the referenced and resulting knowledge graph i.e. SKG and those schemas of datasets.

There are two steps in this experiment. The first thing to do is to determine semantically equivalent etypes between these 3 types of schemas pairwise according to the definitions in section 3. In order to simplify the problem, we assume that the semantic equivalence of etypes is good enough to be symmetric and transitive across schemas. That is to say, if there exist etypes *a*, *b* and *c* from the schema *A*, *B* and *C*, if *a* is semantically equivalent to *b*, and so is *a* to *c*, then *b* and *c* are semantically equivalent. Therefore, we only need to calculate the semantic equivalence of CQ-SKG and Datasets-SKG respectively.

CQ is presented in the form of natural language in the project we utilized. Therefore, we manually extracted the etypes contained in CQ with the help of the project document. The project document provides some lexical alignment, and thanks to that, most of the etypes can be judged to be semantically equivalent on label level. However, there are some exceptions. For example, etype *Company* extracted from CQ and etype *Business_Organization* from SKG are neither semantically close nor synonymous in WordNet[9], but they have same object properties in property level, so they are calculated to be semantically equivalent according to the definitions in section 3. In order to facilitate the next step of our experiment, we standardize all equivalent etypes according to SKG.

As for Datasets, it is a bit more complicated, because the equivalence of etypes contained in different tables of datasets must be determined first before the comparison with SKG. Therefore, individual layer judgement is used. For example, etype *CompanyCategory* and etype *Industry* from 2 different tables have the same individuals and are calculated to be semantically equivalent according to the definitions in section 3. Finally, the comparison with etypes in SKG is the same as that of CQ-SKG.

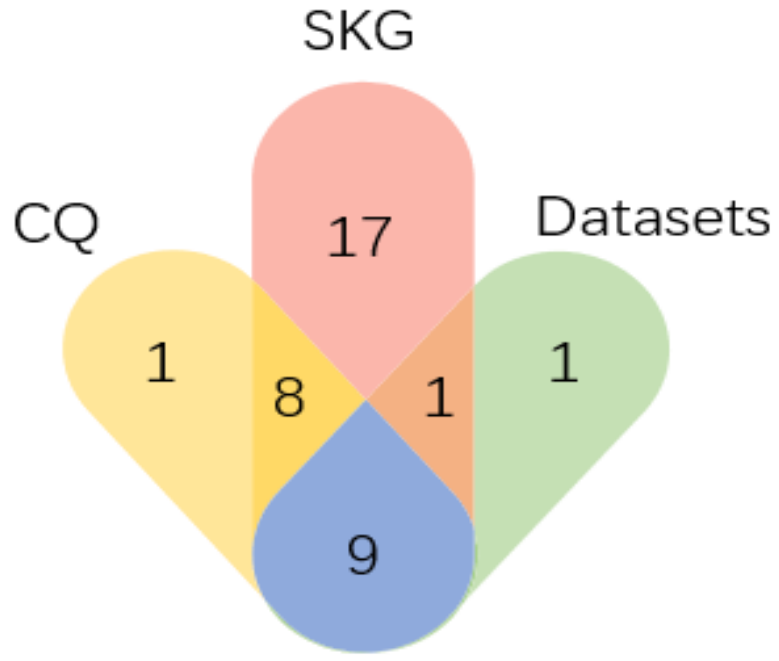

**Figure 7: Equivalent etypes across schemas**

After etype equivalence judgment and standardization, the numeric relationship of SKG, CQ and Datasets in our experiment is shown in Figure 7. The overlapping parts represent equivalent etypes.

The 2nd step of our experiment is to calculate the weights of all etypes in SKG by the method proposed in section 4. We divide these etypes into six importance degrees according to the step size of 0.02 and use them as independent variables of the experiment. See appendix for each weight value of etype and weight values in CQ and Datasets.

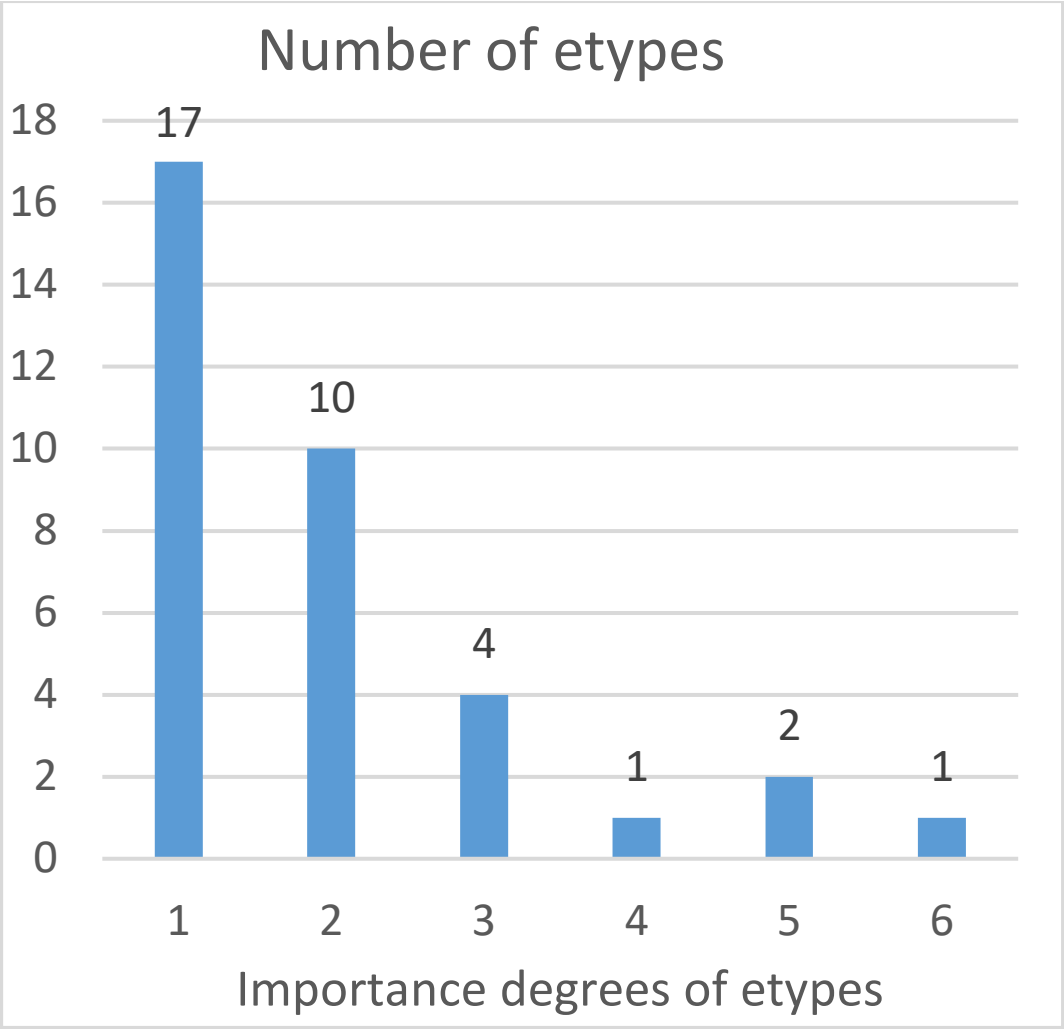

**Figure 8: Number of etypes in each importance degree**

It can be seen from Figure 8 that most etypes in SKG are of low importance. There are only a few core etypes with high weight. 3 calculation methods were involved in out experiment: (1) The method which didn't consider property-based weight calculation. (2) The method which considered weight calculation but didn't involve is-a relationship handling in section 4.2. (3) The method which utilized every metric we proposed in section 4. They will be referred to as Method 1, 2 and 3 in short from now on.

For each method, Coverage and Flexibility of SKG to CQ and SKG to Datasets are calculated ($Cov(A, B)$ indicating the overlap degree of A to B and $Flx(A, B)$ indicating the redundancy degree of A after A covers B). Especially, according to our definition, there is no difference between Method 2 and Method 3 when calculating Coverage, so only Method 2 is used in 5.2.1. During the experiment, one etype belonging to a certain degree of importance in SKG was removed at a time, and the changes of Coverage and Flexibility were observed. Average value was taken when etypes belonging to the same degree of importance were removed.

## 5.2 Results and Analysis

### *5.2.1 Coverage*

As shown in Figure 9, when no etypes are removed, the Coverage value calculated by Method 2 is higher than that calculated by Method 1, because the importance of etypes not included in SKG in CQ is very small; As the importance of removed etypes increases from 1 to 3, the Coverage of Method 2 decreases more and more, because the absence of etypes with high importance in SKG will greatly reduce the Coverage value calculated according to Method 2, while the decline of Method 1 has no obvious change, indicating that the calculation results of Method 2 can better reflect the change of coverage; Especially, when removing etypes with importance degree of 4, the decreasing amplitude of Method 2 becomes smaller, because there is only one etype *Patent* in the importance degree of 4, and the weight of this etype in CQ is not high, so it presents abnormal points as shown in the Figure. When removing etypes with importance degree of 5 and 6, the Coverage value calculated by Method 2 is lower than that calculated by Method 1, because if the most important etypes are not modeled in SKG, the overlap degree of CQ to SKG will decrease greatly, while the calculated value of Method 1 increases instead.

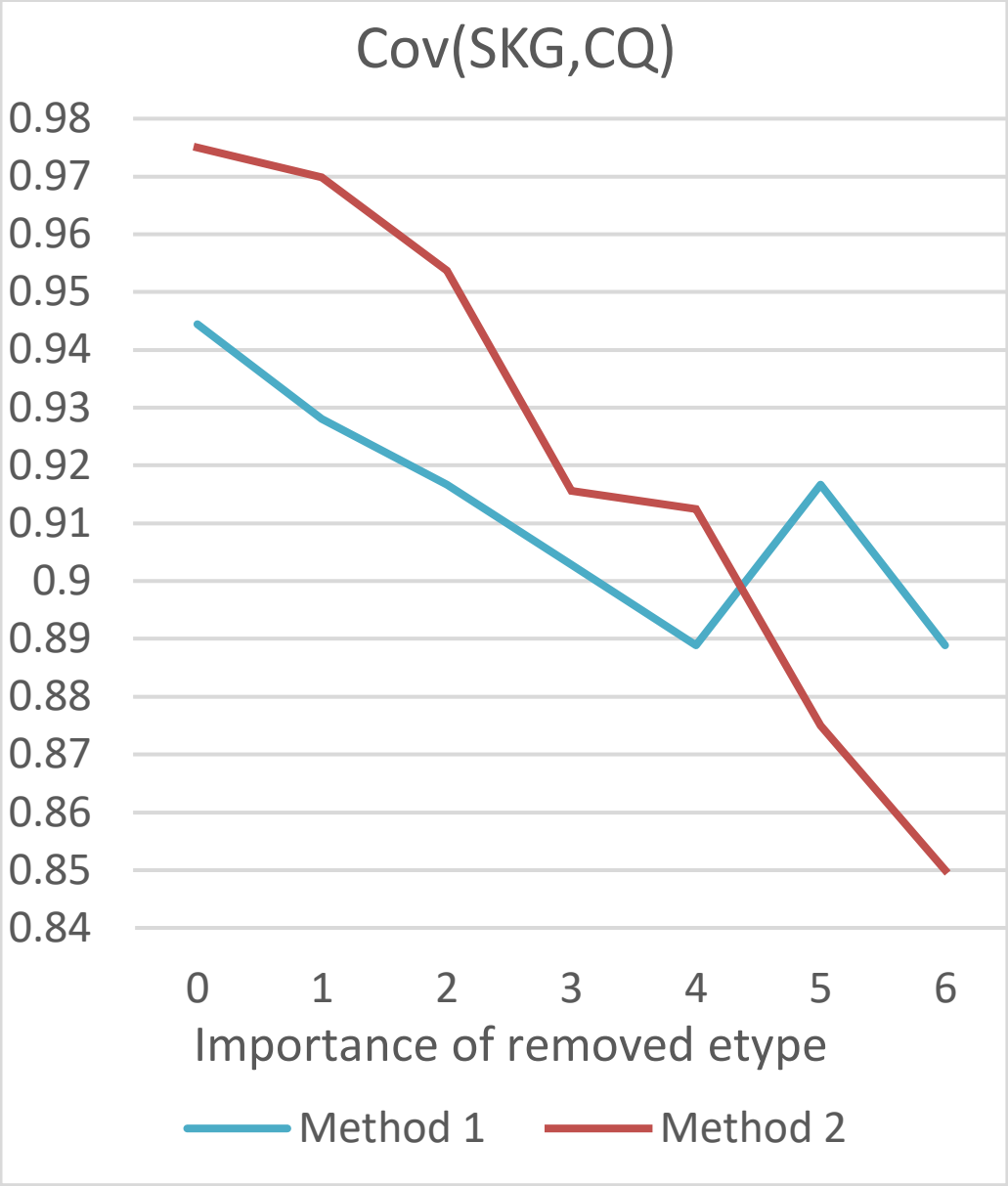

**Figure 9: Coverage of SKG to CQ**

As for Coverage of SKG to Datasets, the initial situation is similar to CQ-SKG as shown in Figure 10, but when removing etype with the importance of 4, the calculated values of the two methods both rise, which is because the etype *Patent* is not included in Datasets of this project, leading to the increase of SKG's overlap degree to Datasets after removing Patent in SKG. When removing etypes with the highest degrees of importance, the calculated value of Method 2 drops significantly, because the weight of the most important etypes in SKG are also very high in Datasets. Removing these etype will lead to a significant decrease of the overlap degree of SKG to Datasets, which is not reflected by the trend of Method 1.

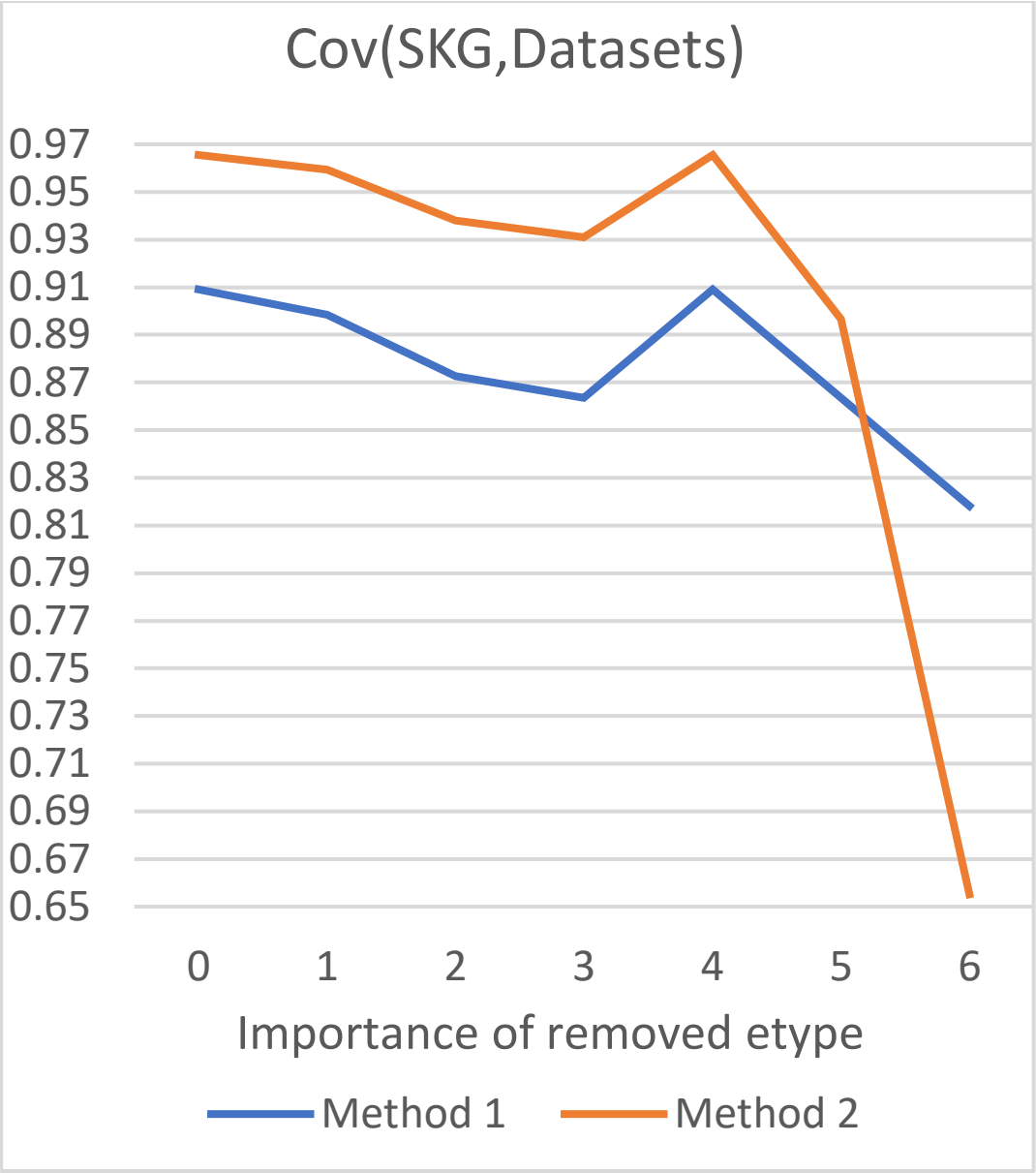

**Figure 10: Coverage of SKG to Datasets**

To sum up, when calculating the Coverage of SKG to CQ and SKG to Datasets, Method 2 which takes weight into consideration is better than Method 1.

*5.2.2 Flexibility*

The higher the importance of an etype in SKG, the more likely it is to appear in CQ, and the greater its weight in CQ, this is consistent with the actual project. At the same time, SKG contains more etypes than CQ. Therefore, when etypes which are important to both CQ and SKG are removed from SKG, the redundancy of SKG to CQ will increase.

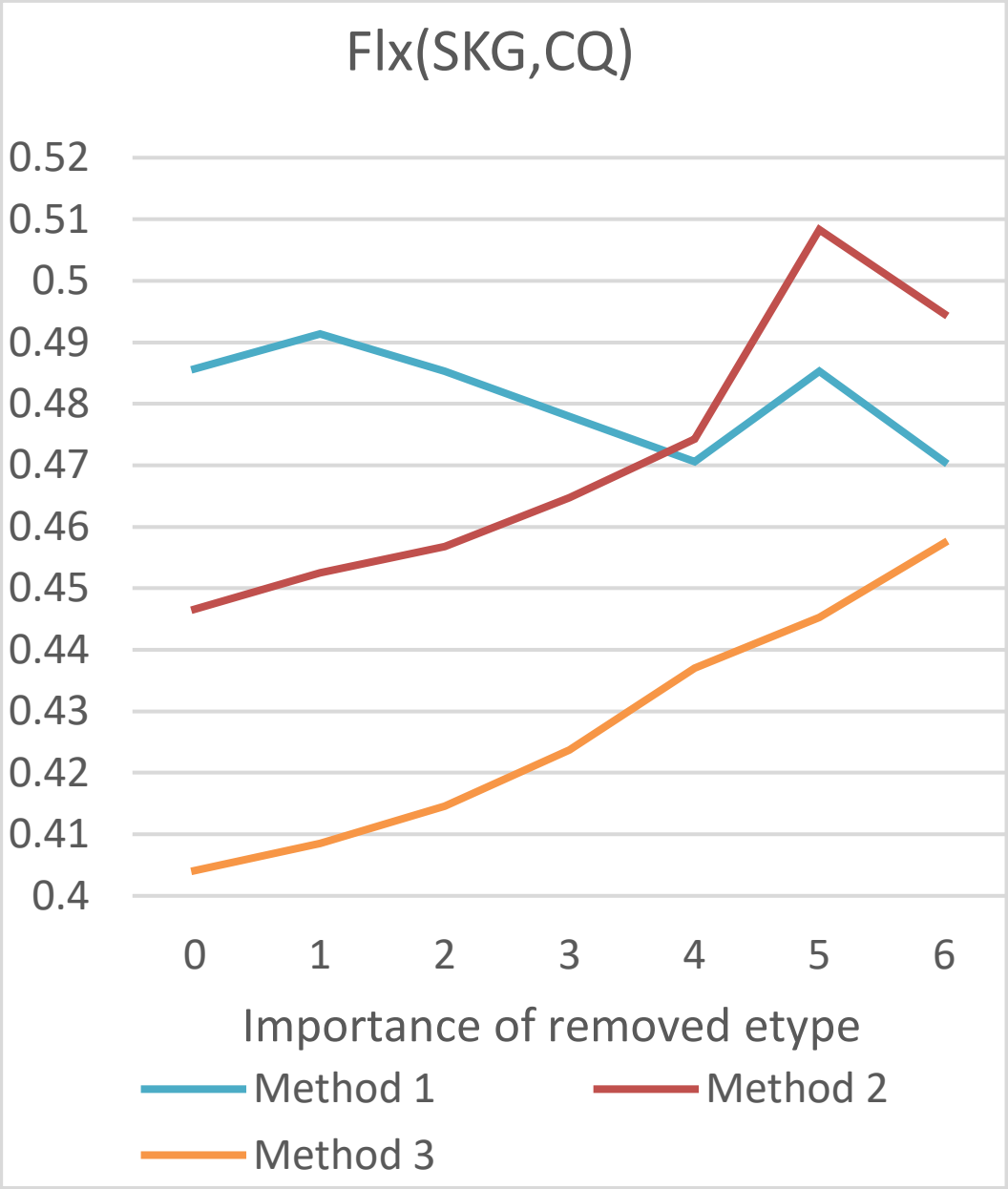

**Figure 11: Flexibility of SKG to CQ**

As shown in Figure 11. Among the three calculation methods, Method 3 is most in line with this trend. The trend of Method 1 is irregular. Although Method 2 also has the trend of increase, the calculated Flexibility values are all greater than Method 3. This is because Method 2 doesn't deal with is-a relationship, and the calculated weight values of etypes which belong to high importance degrees are less prominent, resulting in the calculated redundancy being larger. And it's also the reason why Method 2 has a drop at degree 6.

For Flexibility of SKG to Datasets in Figure 12, the situation is roughly the same as Flexibility of SKG to CQ.

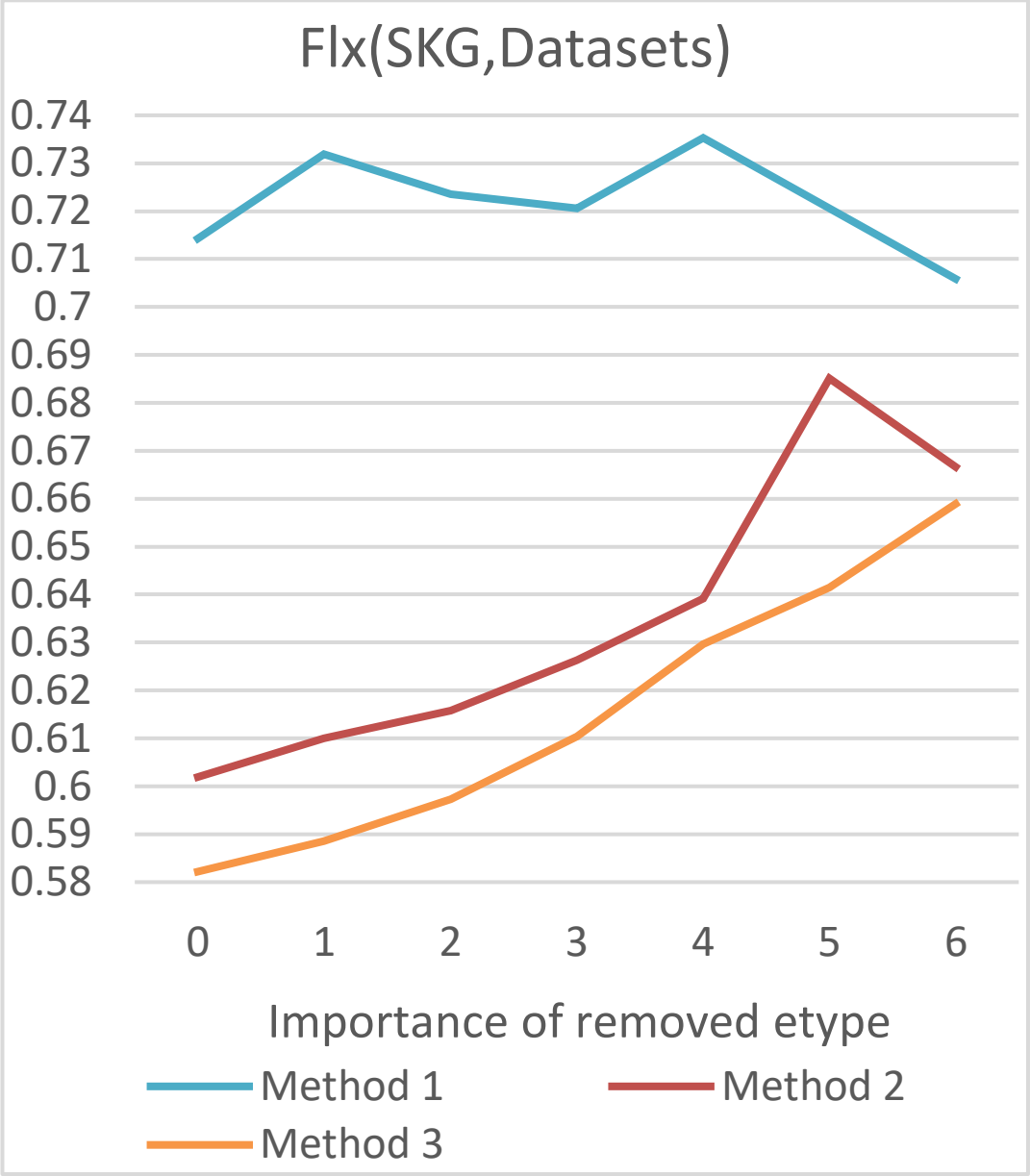

**Figure 12: Flexibility of SKG to Datasets**

To sum up, when calculating the Flexibility of SKG to CQ and SKG to Datasets, Method 3 which utilizes every property-based metrics in section 4 is better than Method 1 and 2.

## 6 State of Art

Data integration is an important problem of big data management. In recent years, many new data integration platforms and concepts have been put forward. Such as Google developed GOODS system [10], Abedjan proposed DataXFormer system [11]. According to the internal DATA CIVILIZER system [12] and the Aurum system [13] which applies the technology of modeling data association based on graph model, etc. In the process of data integration, it is inevitable to encounter low-quality data,Deng et al.[14] put forward SILKMOTH system, which can solve the problem that the attribute values of the same entity in different data sources have certain errors and can not be identified by traditional methods; At the same time, pruning strategy is further introduced to reduce the time complexity of the algorithm. However, this method can not improve the data quality, but also bring about the loss of accuracy. Data quality and data processing efficiency are also concerned[15]. Representative work includes stitching technology proposed by [16], which combines small data into big data and solves the problem of poor effect of small data in data integration. [17] summarizes the development of data integration technologies in recent ten years.

One of the important ways of data integration is to guide the integration at the data level through the schema extracted from the database, which involves ontology reuse[18] and ontology aggregation [19] at the schema level.In the past few years, there are more and more available ontologies, thanks to the help of existing tools, such as Linked Open Verbales (LOV) [20],Bioportal (Biomedicine) [21], etc. However, when aggregating these ontologies across domains, such as TKM[22], etc., we may need to introduce two different ontologies into the third ontology, and then contradictions may arise [23], and the heterogeneous problems [24] that are difficult to deal with in ontology also need to be solved. On the other hand, ontology normalization is very important to realize ontology aggregation.At present, there have been some solutions to the normalization of ontology in the development stage, such as WIDOCO[25], which records the wizard of ontology, and VoCol[26], which supports the integrated environment of version-controlled vocabulary development. Fèrnandez et al. investigated this and found that the heterogeneity between the required conceptualization and the existing ontology concept is an important obstacle to the development of ontology aggregation [27].

Several semi-automatic methods of ontology aggregation have been proposed, such as semi-automatic generation of attribute semantics based on ontology integration [28], automatic generation of new ontology patterns based on ontology reuse[29]. Ontology integration method based on knowledge graph and machine learning technology[19], ontology integration[30] based on ontology matching [31] and its related technologies[32], the model ranking of evaluating ontology based on semantic matching[33], the method of constructing ontology from ontology pattern[34] and compatibility index for comparing and aggregating ontologies[35].However, when these methods evaluate ontologies, they all use qualitative methods to measure the relationship between ontologies and the value of ontologies themselves.The computability of the evaluation method has not been considered.

In 2020, F. Giunchiglia and M. Fumagalli made a preliminary exploration on quantitative evaluation[36], but the evaluation method regarded the importance of all etypes as the same, without considering the differences among different etypes in the same SKG.In the same year, Park tried to measure the importance of nodes in the knowledge graph[37].However,it mainly focused on the influence of inputs from different sources on nodes and did not pay attention to the influence of factors such as the relationship between nodes in the knowledge graph on the importance of nodes.

In solving the problem of data integration and ontology fusion, how to reuse high-quality ontology and data is an important problem, and when distinguishing whether high-quality ontology and data are applicable, the quantitative calculation of measurement index is particularly critical. The important contribution of this paper is to give a reasonable quantitative calculation method of evaluation index in the process of data integration.

## 7 Summary

This paper studies the semantic equivalence of etymology between two SKGs and the evaluation method of overlapping effect between two SKGs based on equivalence. In this paper, the semantic equivalence of etypes is divided into three levels: label, attribute and individual. According to these three levels, a formula to judge whether roots are equivalent is given, which is helpful for the computability of the semantic equivalence method. On the other hand, this paper provides a quantitative evaluation method for measuring the coverage ability and effect between different SKGs. However, in the past, almost all the evaluation methods only stayed in the qualitative analysis stage, and can not be calculated. Therefore, when trying to accurately measure the fusion effect between SKGs, it is very valuable to propose a reasonable quantitative evaluation method. In view of the differences between SKG etymology, this paper proposes a method to measure the importance of etymology by using the object attributes between etymologies, and gives a quantitative calculation method of etymology weights. On this basis, we propose two quantitative calculation formulas of the coverage and flexibility based on etype weight. We have proved the correctness and effectiveness of the method by comparing the results and analysis of a large number of experiments. The results of this study give the judgment method of semantic equivalence and the quantitative calculation and comparison method of two SKGs in the matching process, which is of great significance for realizing the fusion of knowledge and data.